\documentclass[prd,aps]{revtex4-1}

\usepackage{epsfig}
\usepackage{graphicx,color}

\newcommand{\beq}{\begin{equation}}
\newcommand{\eeq}{\end{equation}}
\newcommand{\be}{\begin{eqnarray}}
\newcommand{\ee}{\end{eqnarray}}

\begin{document}

\title{
Early LHC bound on the $W'$ boson mass 
in the nonuniversal gauge interaction model 
}

\author{ Yeong Gyun Kim }

\affiliation{
Department of Science Education,
Gwangju National University of Education, Gwangju 500-703, Korea
}

\author{ Kang Young Lee }
\email{kylee14214@gmail.com}

\affiliation{
Division of Quantum Phases \& Devices,
School of Physics, Konkuk University, Seoul 143-701, Korea
}

\date{\today}

\begin{abstract}

We study the phenomenology of the heavy charged gauge boson 
and obtain the lower bounds on its mass 
with the early LHC data at 7 TeV center-of-mass energy
in the nonuniversal gauge interaction model,
in which the electroweak SU(2) gauge group depends upon the fermion family.
We found that the direct bound on the mass of the $W'$ boson
is compatible to the indirect bound with only the early data of the LHC.

\end{abstract}


\maketitle

The CERN Large Hadron Collider (LHC) has started to operate 
with the center-of-mass (CM) energy of 7 TeV.
The LHC is a discovery machine of the new physics phenomena
beyond the standard model (SM) as well as
a probe of the structure of the electroweak symmetry breaking.
Discovery of a new particle is a clear evidence of the new physics
and the LHC pushes ahead on searching for new particles 
predicted in various models beyond the SM, 
e. g. supersymmetric particles, 
exotic Higgs bosons, the Kaluza-Klein states, 
and extra gauge bosons etc..

Recently the CMS \cite{cms} and ATLAS \cite{atlas} collaborations 
has reported the search results for the extra charged gauged boson, $W'$,
through the leptonic decay channels
with the early data of 36 pb$^{-1}$ collected in 2010 at the LHC.
From the absence of the excess above the SM expectations 
in the transverse mass distribution of a lepton,
the mass bound of $W'$ boson is obtained to be about 1.4 TeV,
assuming the $W'$ boson couplings are same as 
those of the ordinary $W$ boson.
The $W'$ boson is predicted in many new physics models 
such as the left-right symmetric model \cite{lr},
extra dimensional models \cite{extraD}, Little Higgs models \cite{little}
and models with extended gauge symmetry \cite{topflavor,331}.

We consider an extension of the SM with a separate SU(2) group
which acts only on the third generation while the first two generations
couple to the usual SU(2) group \cite{topflavor}.
The phenomenology of this model has been intensively studied 
in the literatures, 
using the electroweak precision test with $Z$-pole data 
and the low-energy data 
\cite{topflavor,lee1,lee2,malkawi2}.
The gauge group of this model arises as a theory
at an intermediate scale in the path of gauge
symmetry breaking of noncommuting extended technicolor models \cite{ETC}.
In this model, 
the SU(2) gauge coupling constants depend upon the fermion family 
and are nonuniversal in general.
The nonuniversality of the gauge couplings leads to
the exotic phenomena in the charged currents and neutral currents interactions
although they are suppressed by the high energy scale of new physics.
In the charged currents interactions,
the unitarity of the Cabibbo-Kobayashi-Maskawa (CKM) matrix 
is violated explicitly.
In the neutral currents interactions,
the flavour-changing neutral current (FCNC) interactions
arise at tree level. 
The unitarity violation of the CKM matrix and 
the lepton flavor violating processes induced by the FCNC 
strongly constrain the model parameters \cite{ckm,lfv}.

The $W'$ boson exists in our model, 
since we have an additional SU(2) gauge symmetry 
of which mass is of order the new physics scale.
In this work, we study the $W'$ boson with the early LHC data
collected at the first run of the LHC.
We obtain constraints on the model parameters 
of the $W'$ mass and the mixing angle between two SU(2) groups
from the lack of the signal of $W'$ boson at the LHC
and will show that the direct bound from the early LHC data
is compatible to the indirect bound from the unitarity of the CKM matrix.

We consider the new physics model with the electroweak gauge group 
$SU(2)_l \times SU(2)_h \times U(1)_Y $
where the first and the second generations couple to $SU(2)_l$ 
and the third generation couples to $SU(2)_h$.
We write the covariant derivative as
\begin{equation}
 D^{\mu} = \partial^{\mu} + i g_l T^{a}_{l} W^{\mu}_{l,a}
          + i g_h T^{a}_{h} W^{\mu}_{h,a}
          + i g^{\prime} \frac{Y}{2} B^{\mu} ,
\end{equation}
where $ T^{a}_{l,h}$ denotes the $SU(2)_{(l,h)}$ generators and
$Y$ the $U(1)$ hypercharge. 
The electric charge is defined as $ Q = T_{3l} + T_{3h} + Y/2 $.
The left-handed quarks and leptons, $Q^{1,2}_L$ and $L^{1,2}_L$,
of the first and second generations
transform as $(2,1,1/3)$, $(2,1,-1)$
and those in the third generation, $Q^3_L$ and $L^3_L$ 
as $(1,2,1/3)$, $(1,2,-1)$
under $SU(2)_l \times SU(2)_h \times U(1)_Y $
while right-handed quarks and leptons transform as (1,1,2$Q$).

We introduce an additional bidoublet scalar field $\Sigma$,
transforming as (2,2,0),
to break the $ SU(2)_l \times SU(2)_h \times U(1)_Y $ gauge symmetry
into the $ SU(2)_{l+h} \times U(1)_Y $,
of which vacuum expectation values (VEV) is given by
\be
 \langle \Sigma \rangle = \left( \begin{array}{cc}
             u&0 \\
             0&u
           \end{array} \right).
\ee
The electroweak symmetry breaking arises
at the electroweak scale $v$ 
by the VEV of the (2,1,1) scalar field $\Phi$,
which is corresponding to the SM Higgs boson.
We require that the scale $u$ is
higher than the electroweak scale $v$
and introduce the small parameter $ \lambda \equiv v^2/u^2 $.
Note that the third generation fermions do not couple to $\Phi$
and they should get masses by other way, e.g introducing
higher dimensional operators or another Higgs doublet etc..
The different mechanism of mass generation
could be the origin of the heavy masses of the third generation.
We do not discuss the details of the Higgs sector in this paper.
The Higgs sector of this model has been discussed 
in Ref. \cite{chiang}.

After the symmetry breaking,
the gauge coupling constants are parametrized by
\beq
g_l \sin \theta \cos \phi = g_h \sin \theta \sin \phi
= g^{\prime} \cos \theta = e
\eeq
in terms of the electromagnetic coupling $e$, 
the weak mixing angle $\theta $ and the new mixing angle $\phi$ 
between $SU(2)_l$ and $SU(2)_h$.
We demand that all of the gauge couplings are perturbative
so that $g_{(l,h)}^2/4 \pi < 1$,
which results in $ 0.03 < \sin^2 \phi < 0.96 $.

We have additional $W'$ and $Z'$ gauge boson with masses 
\be
m_{W'}^2 = m_{Z'}^2
= \frac{m_0^2}{\lambda \sin^2 \phi \cos^2 \phi} ,
\ee
where $m_0 = ev/(2 \sin \theta)$ is 
the ordinary $W$ boson mass in the leading order.
We find that the $W'$ and $Z'$ masses are degenerate in this model.
The charged current interactions for $W'$ boson is given by
\be
{\cal L}_{CC} = V_{UD} \bar{U}_L \gamma_\mu {G'}_L {W'}^\mu D_L + {\rm H.~c.},
\ee
where $U_L=(u_L, c_L, t_L)^T$, $D_L =(d_L, s_L, b_L)^T$
and the couplings are
\be
{G'}_L &=& -\frac{g}{\sqrt{2}} \tan \phi,
            ~~~~~~~~~~~~~~~~~~~~~{\rm 1st,~2nd~~ generations},
\nonumber \\
     &=& -\frac{g}{\sqrt{2}} \tan \phi \left(1-\frac{1}{\sin^2 \phi} \right),
                    ~~~~~{\rm 3rd~~ generations}.
\ee

The tree level decay rates of $W'$ boson are 
obtained by the replacements of the couplings and mass
of $W$ boson by those of $W'$ in the SM decay rates,
given by
\be
\Gamma (W' \to f \bar{f}') = \Gamma_0 \frac{m_{W'}}{m_W} \cdot \tan^2 \phi,
\ee
for the first and second generations and
\be
\Gamma (W' \to f \bar{f}') 
        = \Gamma_0 \frac{m_{W'}}{m_W} 
     \cdot \tan^2 \phi \left(1-\frac{1}{\sin^2 \phi} \right)^2,
\ee
for the third generations,
where $\Gamma_0 = \Gamma(W \to e^- e^+)$ in the SM.
We ignore the final state masses 
except for the decay involving top quark,
since the $W'$ mass is more than 600 GeV in this analysis.

\begin{figure}[t!]
\centering
\includegraphics[height=12cm]{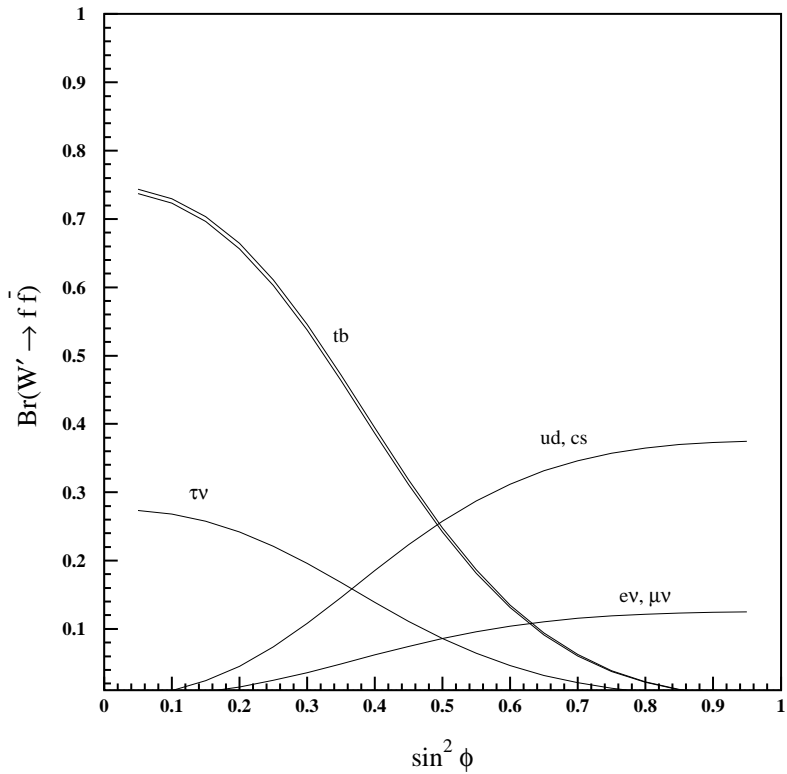}
\caption{
Branching ratios of the $W'$ boson with respect to $\sin^2 \phi$.
}
\end{figure}

The branching ratios of $W'$ boson are depicted in Fig. 1.
Since the $W'$ mass is an overall factor for the decay width
except for the decay into top quark,
the branching ratios depend only on $\sin^2 \phi$.
Only the $W' \to tb$ decay shows a small splitting due to
the top quark mass effects, one is for $m_{W'}=600$ GeV
and the other for $m_{W'}=2$ TeV.
Such splittings for other fermions are negligible.
We see that decays into the third generations are dominant
in the small $\sin^2 \phi$ region.
Since the angle $\phi$ represents the mixing 
between $SU(2)_l$ and $SU(2)_h$,
$W'$ boson is almost $W_h$ boson and coupled to the third generations
in the small $\phi$ limit.
The mixing is maximal if $\sin^2 \phi \to 1$,
then $W'$ is almost $W_l$ and decays dominantly
into the first and second generations.

\begin{figure}[ht]
\centering
\includegraphics[height=13cm]{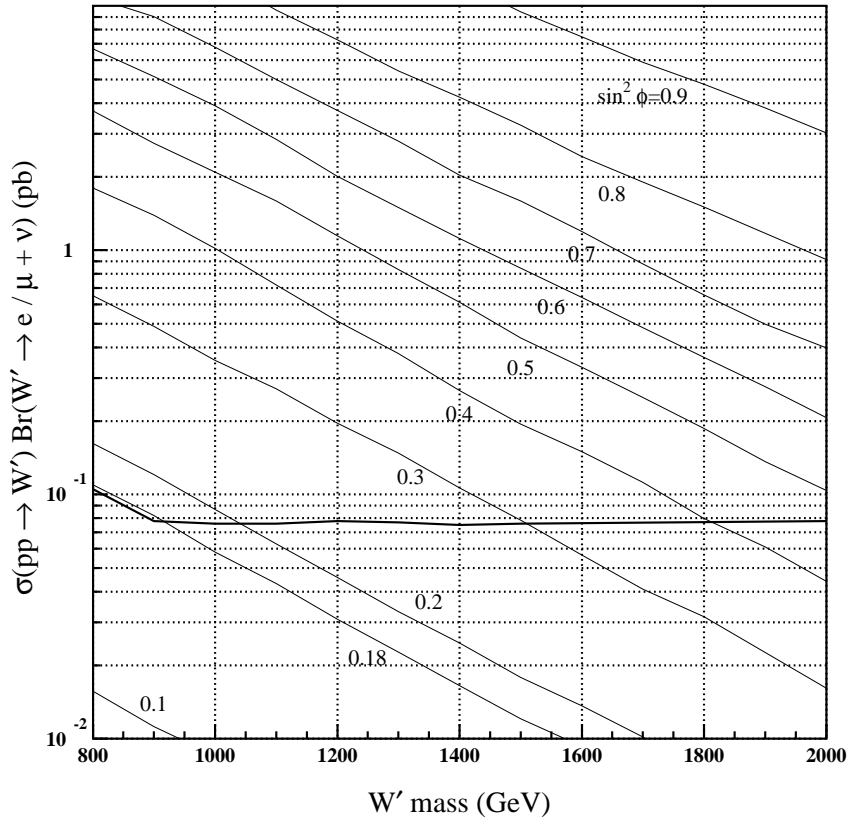}
\caption{
Cross sections multiplicated by branching ratios into leptons 
in $pp \to q \bar{q}' \to W' \to l \nu$ processes.
The thick line is the early LHC bound obtained by the CMS group
\cite{cms}.
}
\end{figure}

We consider the single production of $W'$ boson 
in the $pp$ collisions at the LHC.
Using about 36 pb$^{-1}$ data collected at the LHC in 2010,
the CMS and the ATLAS groups have searched for $W'$ boson
through the transverse mass distributions
in $W' \to e \nu / \mu \nu$ decays
and determine the upper limit on the cross section
from the absence of the $W'$ signal in the early LHC data.

We calculate the production cross sections
in terms of $m_{W'}$ and $\sin^2 \phi$ in our model
by using PYTHIA 6.4 \cite{PYTHIA}.
Our results are shown in Fig. 2 
together with the limit from the LHC data at the 95 \% C. L., 
presented as a thick line.
Here we use the bound of the CMS collaboration 
combining the decays into electron and muon \cite{cms}.
The region above the thick line is excluded.
Comparing the cross sections of our model with the limit
from the early LHC data,
we determine the bound on the $W'$ mass with respect to $\sin^2 \phi$.
For instance, the lower bound of $m_{W'}$ is 1.5 TeV
when $\sin^2 \phi=0.3$.

In the previous analysis in Ref. \cite{topflavor,lee1,malkawi2}
with the LEP and SLC data,
the atomic parity violation (APV), 
the low-energy neutral currents interaction data,
the indirect constraints on the model parameters 
($\sin^2 \phi, m_{W'}$) has been provided.
The constraint from the electroweak precision test 
with the data at the $Z$-pole
is stronger than those of the low-energy experiments.

More detailed phenomenology on this model has been studied
to improve the indirect constraints
\cite{ckm,lfv}.
The nonuniversality of the SU(2) couplings
derives modifications on both the charged current 
and the neutral current interactions.
For the quark sector,
the charged current couplings are measured
by elements of the CKM matrix which is unitary.
However in this model, 
the CKM matrix is no more unitary 
due to the nonuniversality of the gauge coupling. 
Moreover, additional charged current interactions
via the $W'$ boson exist in this model.
Thus the observed CKM matrix is the combination of
the $W$ boson and $W'$ boson exchanges.
We define the observed CKM matrix
in the low-energy four fermion effective Hamiltonian
for the semileptonic quark decay
and extract $V_{CKM}$ in this model;
\be
V_{CKM} = V_{CKM}^0 + \epsilon^c {V_U}^\dagger M V_D
              + \left( \frac{G'^c_L}{G^c_L} \right)^2
                \frac{m_W^2}{m_{W'}^2}
     \left( V_{CKM}^0 + \epsilon'^c {V_U}^\dagger M V_D \right)
\ee
where the suppression terms
$\epsilon^c = \lambda \sin^2 \phi + {\cal O}(\lambda^2)$
and $\epsilon'^c = 1/\sin^2 \phi + {\cal O}(\lambda)$.
The $3 \times 3$ matrices
$V_U$ and $V_D$ are unitary matrices
which diagonalize up- and down-type quarks,
$V_{CKM}^0 \equiv {V_U}^\dagger V_D$ is the CKM matrix 
defined in the SM,
and $M \equiv \delta_{3i} \delta_{3j}$ are defined 
to express the nonuniversal terms.
We simplify the expression to obtain
$V_{CKM}=V_{CKM}^0 (1 + \lambda \sin^4 \phi)$.

\begin{figure}[t!]
\centering
\includegraphics[height=12cm]{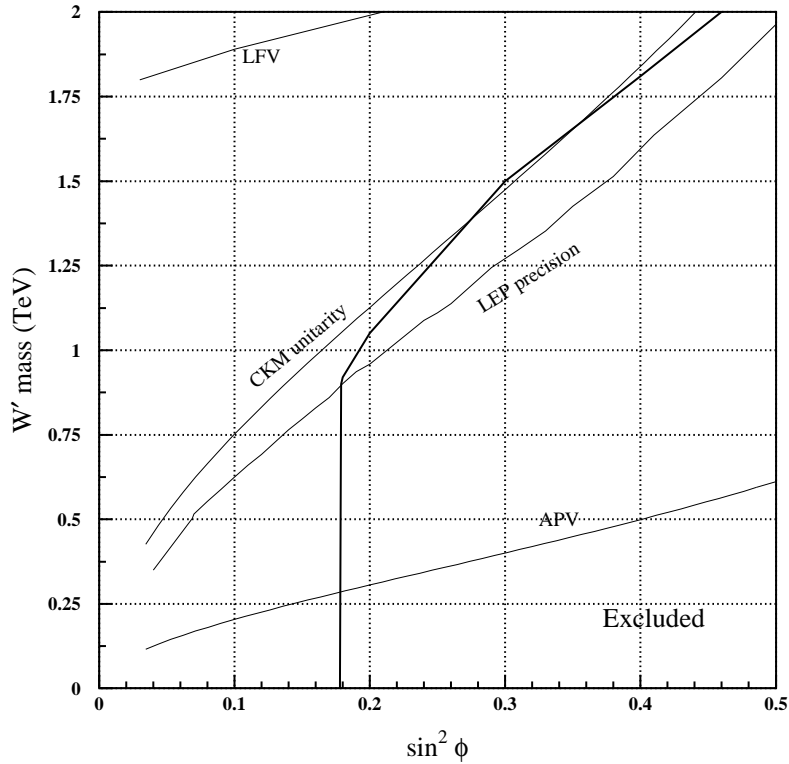}
\caption{
Allowed parameters on $(\sin^2 \phi, m_{W'})$ space
with direct constraints from the early LHC data and indirect constraints
from various experiments.
The thick line is the direct bounds from the LHC data.
}
\end{figure}

In order to measure the unitarity of the CKM matrix,
the unitarity violating term $\Delta$ is defined by
$|V_{ud}|^2 + |V_{us}|^2 + |V_{ub}|^2 = 1-\Delta$,
which is measured to be $\Delta = 0.0009 \pm 0.0010$
from the nuclear beta decays, kaon decays and $B$ decays
\cite{czarnecki}.
Since we derive $\Delta = 2 \lambda \sin^4 \phi$ in our model,
we have constraints on the parameter space ($\sin^2 \phi, m_{Z'}$),
which is the stronger limit than those of the electroweak precision test 
and the low-energy neutral current data \cite{ckm}.

If the SU(2) couplings depend on the fermion family,
the neutral currents are not simultaneously diagonalized 
with the fermion mass marix
and the FCNC interactions generically arise in our model.
In the lepton sector, the FCNC interactions 
lead to dangerous lepton flavour violating (LFV) processes at tree level.
The LFV processes have not been observed in the experiments so far
and is bounde very strongly.
Although the FCNC contain additional unidetermined parameters,
we can obtain the conservative constraints on the model parameters 
which are less sensitive to the assumptions on neutrino masses
\cite{lfv}.

Finally, we obtain the direct bounds on $W'$ masses
on $(\sin^2 \phi, m_{W'})$ from Fig. 2
and show them together with all the indirect constraints in Fig. 3.
We find that the direct bounds from the early LHC data
is compatible to the constraints from the CKM unitarity
if $\sin^2 \phi > 0.18$.
Note that both of the production and decay of $W'$ boson in
of $p p \to u d \to W^\pm \to l \nu$ process becomes small 
in the small $\sin^2 \phi$ region
and the constraints are very weak.

In conclusion, we obtain the direct bound on the mass of the $W'$ boson 
with the early LHC data 
in the nonuniversal $SU(2)_l \times SU(2)_h \times U(1)_Y$ model.
Since the LHC have already collected more data from the 2011 run
than total data collected in 2010,
the direct bound with the LHC data will be improved soon.

\acknowledgments
YGK is supported by the Basic Science Research Program through 
the National Research Foundation
of Korea (NRF) funded by the Korean Ministry of
Education, Science and Technology (2010-0010312).
KYL is supported in part by WCU program through the KOSEF funded
by the MEST (R31-2008-000-10057-0)
and the Basic Science Research Program through the National Research Foundation
of Korea (NRF) funded by the Korean Ministry of
Education, Science and Technology (2010-0010916).

\def\PRD #1 #2 #3 {Phys. Rev. D {\bf#1},\ #2 (#3)}
\def\PRL #1 #2 #3 {Phys. Rev. Lett. {\bf#1},\ #2 (#3)}
\def\PLB #1 #2 #3 {Phys. Lett. B {\bf#1},\ #2 (#3)}
\def\NPB #1 #2 #3 {Nucl. Phys. B {\bf #1},\ #2 (#3)}
\def\ZPC #1 #2 #3 {Z. Phys. C {\bf#1},\ #2 (#3)}
\def\EPJ #1 #2 #3 {Euro. Phys. J. C {\bf#1},\ #2 (#3)}
\def\JHEP #1 #2 #3 {JHEP {\bf#1},\ #2 (#3)}
\def\IJMP #1 #2 #3 {Int. J. Mod. Phys. A {\bf#1},\ #2 (#3)}
\def\MPL #1 #2 #3 {Mod. Phys. Lett. A {\bf#1},\ #2 (#3)}
\def\PTP #1 #2 #3 {Prog. Theor. Phys. {\bf#1},\ #2 (#3)}
\def\PR #1 #2 #3 {Phys. Rep. {\bf#1},\ #2 (#3)}
\def\RMP #1 #2 #3 {Rev. Mod. Phys. {\bf#1},\ #2 (#3)}
\def\PRold #1 #2 #3 {Phys. Rev. {\bf#1},\ #2 (#3)}
\def\IBID #1 #2 #3 {{\it ibid.} {\bf#1},\ #2 (#3)}

\end{document}